\documentclass[hyper]{JHEP3}

\input{epsf}
\usepackage{epsfig}
\usepackage{amssymb}
\usepackage{amsfonts}
\usepackage{amsbsy}
\usepackage[all]{xy}
\usepackage{amsmath}

\usepackage{amssymb,amscd}
\usepackage{mathrsfs}
\usepackage{amsmath,amsthm}
\def\be{ \begin{equation} }
\newcommand{\bel}[1]{\begin{equation}\label{#1}}
\def\ee{ \end{equation}}

\def\dim{{\rm dim}}

\def\exp{{\rm exp}}

\def\I{{\rm i}}

\def\ker{{\rm ker}}

\def\mod{{\rm mod}}

\def\Tr{{\rm Tr}}

\def\half{\frac{1}{2}}
\def\p{\partial}

\def\one{{\hbox{ 1\kern-.8mm l}}}

\def\CF {{\cal F}}

\def\CL {{\cal L}}
\def\CM {{\cal M}}

\def\CP {{\cal P}}
\def\CR {{\cal R}}

\def\CW {{\cal W}}

\def\CQ {{\cal Q}}
\def\CS {{\cal S}}

\def\CU {{\cal U}}

\def\IH{\mathbb{H}}

\def\IR{{\mathbb{R}}}

\def\IZ{{\mathbb{Z}}}

\def\fb{\mathfrak{b}}
\def\fB{\mathfrak{B}}

\def\fg{\mathfrak{g}}
\def\fG{\mathfrak{G}}

\def\fu{\mathfrak{u}}

\def\rmk#1{\bigskip\noindent{\bf Remarks} }

\title{A Brief Summary Of Global Anomaly Cancellation In Six-Dimensional Supergravity}

\author{Samuel Monnier$^1$ and Gregory W.~Moore$^2$ \\
$^1$ Section de Math\'ematiques, Universit\'e de Gen\`eve \\
2-4 rue du Li\`evre, 1211 Gen\`eve 4, Switzerland\\
$^2$ NHETC and
$~~$Department of Physics and Astronomy, Rutgers University \\
$~~$126 Frelinghuysen Rd., Piscataway NJ 08855, USA\\
\\
{\tt samuel.monnier@gmail.com, gmoore@physics.rutgers.edu } }

\abstract{This is a short summary of a talk at Strings 2018. v1: August 3, 2018. 
v2:    \today
}

\begin{document}

\section{Introduction}

This paper is a brief summary of a talk at Strings 2018 in Okinawa
\cite{TalkURL}.
It is also an informal and less-technical summary of our paper \cite{MonnierMoore}.

The motivation for the present work is the general question of the
relation of apparently consistent low energy theories of quantum
gravity to string theory: We would like to find new consistency
conditions to restrict the possible low energy effective supergravity
theories and at the same time find new string constructions. The
interesting question is whether every consistent low energy supergravity
theory can be obtained from a string-theory construction. The
state of the art in this general subject is summarized in \cite{Brennan:2017rbf,TaylorTASI}.
One way to impose consistency conditions
on low energy supergravity theories is via anomaly cancellation. In the case of
six-dimensional supergravity, while the Green-Schwarz mechanism was
described some time ago \cite{Green:1984bx,Sadov:1996zm,Sagnotti:1992qw}, only
recently have the anomaly cancellation conditions been systematically investigated.
See \cite{TaylorTASI} for a review. Moreover, while previous
investigations have included some considerations of global anomaly cancellation,
no completely systematic account of global anomalies in this context has yet been given.
It is the purpose of \cite{Monnier:2017oqd,MonnierMoore} to begin to fill this gap.

The papers \cite{Monnier:2017oqd,MonnierMoore} contain three main results:
First, a necessary condition for anomaly cancellation is stated which summarizes
and extends all previous results.
\footnote{More accurately, \cite{Monnier:2017oqd} uses both
 global anomaly cancellation and tadpole cancellation of
string charge. It is possible the tadpole cancellation conditions can
be rederived from global anomaly cancellation on the worldsheets of the
six-dimensional strings, (somewhat in the spirit of the work of
Polchinski and Cai \cite{Polchinski:1987tu}) but that is beyond the scope of the present work. }
 Second, a necessary and sufficient condition
is given for global anomaly cancellation in terms of triviality of a certain
7-dimensional spin topological field theory (defined below). Third, the
necessary conditions of \cite{Monnier:2017oqd} were verified (in that same paper) to hold in the
case of F-theory compactifications. A novel aspect of this third result is that one
must know the global structure of the connected component of the gauge group for the
vectormultiplets. In this summary we will focus on just the  first two of these
three results.

\section{Data For Six-Dimensional Supergravity}

The field content of six-dimensional $(1,0)$ supergravity is determined by
giving three pieces of data: First, a compact Lie group $G$ for the vectormultiplets,
second, a quaternionic representation of $G$, denoted $\CR$, for the half-hypermultiplets, and third
an integral lattice $\Lambda$ of Lorentzian signature $(1,T)$ defining the tensormultiplets.
The lattice may be regarded as a lattice of string charges for strings charged under the 2-form
gauge potentials in the tensormultiplets. Of course, in addition to this, we add the supergravity multiplet.
Given this data the equations of motion of the classical supergravity theory are uniquely determined
\cite{Riccioni:2001bg}.

In the quantum theory the chiral fermions and the (anti-) self-dual tensor fields have gauge and
gravitational anomalies \cite{AlvarezGaume:1983ig}. The Green-Schwarz mechanism for the cancellation of the perturbative anomalies
is well known  \cite{Green:1984bx,Green:1987sp,Polchinski:1998rr,Sadov:1996zm,Sagnotti:1992qw}: We compute the anomaly $8$-form from the data $(G,\CR, \Lambda)$:
\be\label{eq:1}
\begin{split}
I_8 & =\half  \left[ Tr_{\CR} e^{\frac{F}{2\pi\I}}  \hat A + \cdots \right]_8 \\
& \sim   (\dim_{\IH} \CR - \dim_{\IR} G + 29 T - 273) \Tr(R^4) + (9-T) (\Tr R^2)^2 + \cdots\\
\end{split}
\ee
(some complicated numerical coefficients are suppressed in the second line). Then we seek to factorize it
\be\label{eq:2}
I_8 = \half Y^2
\ee
where $Y \in \Omega^4(\CW; \Lambda_{\IR} ) $ is a $4$-form on some (as yet unspecified) 8-manifold $\CW$
with values in the vector space $\Lambda_{\IR}:=\Lambda\otimes \IR$. Note that this vector space inherits
a Lorentzian signature quadratic form from $\Lambda$ and in \eqref{eq:2} we implicitly use that form to
contract the two factors of $Y$. This convention will be used henceforth without comment.
Next, we interpret $Y$ as a background magnetic current for the tensormultiplets and modify the Bianchi
identity for the fieldstrength $H\in \Omega^3(\CM_6;\Lambda_{\IR})$ to $dH = Y$. Here $\CM_6$ is the
(compact, Euclidean) spacetime of the six-dimensional supergravity. As is well-known, this forces the
2-form gauge potentials to shift under diffeomorphisms and to transform under vectormultiplet gauge
transformations. We now merely add the counterterm to the exponentiated action $e^{-S}$ in the
supergravity path integral:
\bel{eq:3}
e^{-S} \rightarrow e^{- S} e^{-2\pi\I \half \int_{\CM_6} B Y } \, .
\ee

It is worth noting that there are a few difficulties with this textbook story: First, in topologically
nontrivial situations $B$ is not globally defined, so the six-form $BY\in \Omega^6(\CM_6)$ is not globally defined.
A standard way to deal with this is to extend $B$ and $Y$ to forms on a seven-dimensional manifold
$\CU_7$ so that $\p \CU_7 = \CM_6$. Then, since $dB$ is better-defined than $B$ we can attempt to
substitute
\bel{eq:4}
\half \int_{\CM_6} B Y  \rightarrow \half \int_{\CU_7} dB Y
\ee
Of course this only makes sense if the result is independent of extension. In fact, it is not.
Even the \underline{difference} of Green-Schwarz terms  $\half \int_{\CU_7} (H_1 - H_2) Y$
for two tensormultiplet fields in the same gauge and gravitational background is ill-defined if we
use the standard quantization condition on the periods: $[H_1-H_2]\in H^3_{dR}(\CM_6; \Lambda)$. The problem here is that
the factor of $1/2$ leads to a sign ambiguity. These considerations motivate our more precise
mathematical discussion of the Green-Schwarz term given below.

In order to proceed it is useful to decompose the Lie algebra $\fg$ of $G$ as
a sum over simple Lie algebras $\fg_i$ and a basis of $\fu(1)$ factors:
\bel{eq:5}
\fg = \oplus_i \fg_i \oplus_I \fu(1)_I \, .
\ee
Then the general form for $Y$ is
\bel{eq:6}
Y = \frac{a}{4} p_1 - \sum_i b_i c_2^i + \half \sum_{I,J} b_{IJ} c_1^I c_1^J
\ee
where $p_1$, $c_2^i$, and $c_1^I$ are the Chern-Weil representatives of the indicated
cohomology classes provided by the metric and vectormultiplet gauge fields:
\bel{eq:7}
p_1 := \frac{1}{8\pi^2} \Tr_{\rm vec} R^2 \qquad \qquad c_2^i := \frac{1}{16 \pi^2 h_i^\vee} \Tr_{\rm adj} F_i^2 \qquad \qquad
c_1^I := \frac{1}{2\pi} F_I
\ee
Here $h_i^\vee$ is the dual Coxeter number of $\fg_i$.
The vectors $a, b_i, b_{IJ} \in \Lambda_{\IR}$ appearing in \eqref{eq:6} are, by definition, the \emph{anomaly coefficients}.
(We will see below that there are, in addition, new anomaly coefficients associated with torsion classes which have not
been discussed before.)

We note that just the very existence of a factorization such as \eqref{eq:2} puts nontrivial constraints on
the data $(G,\CR, \Lambda)$. For example, we have the famous constraint
\bel{eq:8}
\dim_{\IH} \CR - \dim_{\IR} G + 29 T - 273 = 0 \, .
\ee
Moreover, the existence of a factorization strongly constrains the anomaly coefficients.
For example, we must have $a^2 = 9-T$. The full set of constraints for factorization is
expressed in terms of group-theoretical factors associated with $G$ and $\CR$. See, for example,
equations $(2.19)$ and $(2.20)$   in \cite{Monnier:2017oqd} for the full list of conditions.
These constraints have been extensively analyzed in the
literature.

Even when a factorization \eqref{eq:2} exists it is not unique. Therefore, the full data needed to
define a perturbatively consistent supergravity theory is the set of data $(G,\CR,\Lambda, a,b_i,b_{IJ})$.
The global anomaly conditions found in
\cite{Monnier:2017oqd} are best stated as follows: First, we must have $a \in \Lambda$.
Next, the data $(b_i, b_{IJ})$ should be regarded as defining a quadratic form on $\fg$ valued in $\Lambda_{\IR}$.
After all, we use a symmetric invariant bilinear form to define the Chern-Weil representatives
in \eqref{eq:7}. We denote the corresponding quadratic form by $\bar b$.
It turns out that the space of such bilinear forms has an interpretation in
topology. It is naturally isomorphic to the cohomology group $\CQ:=H^4(BG_1; \Lambda_{\IR})$, where
$G_1$ is the connected component of the identity in $G$. Inside the vector space $\CQ$ there
is a lattice $H^4(BG_1;\Lambda) \subset \CQ$. The strongest necessary condition in \cite{Monnier:2017oqd}
states that
\bel{eq:9}
\bar b \in 2 H^4(BG_1;\Lambda).
\ee
The  derivation in \cite{Monnier:2017oqd} proceeds by requiring
that the supergravity theory can be put on an arbitrary spin 6-manifold with arbitrary $G$-bundle.
Then, cancellation of string charge in the compact Euclidean spacetime implies that for all
$\Sigma \in H_4(\CM_6;\IZ)$ we must have $\int_{\Sigma} Y \in \Lambda$ in order to cancel
the background string charge with strings.
\footnote{By contrast, arguments in \cite{Monnier:2017oqd}, similar to those in
\cite{KMT2} and other references that are based on   manipulations of Casimirs in group theory, together
with considerations of certain global anomalies lead to weaker conditions, namely conditions 1-8 in
Section 2.3 of \cite{Monnier:2017oqd}. The condition \eqref{eq:9} is rederived in \cite{MonnierMoore}
from consistency of the construction of the Green-Schwarz term. We note that the weaker
condition
$\bar b\in 2 H^4(B\widetilde{G_1};\Lambda)$ where $\widetilde{G_1}$ takes the universal
cover of the semisimple part of $G_1$ can be stated more concretely as
the condition that $b_i, \half b_{II}, b_{IJ} \in \Lambda$. Another way of stating
$\bar b\in 2 H^4(B\widetilde{G_1};\Lambda)$ is that $\bar b$ is an even $\Lambda$-valued quadratic form on the coroot
lattice of $G$, while \eqref{eq:9} requires it to be an even $\Lambda$-valued form on
the cocharacter lattice of $G$.   }
The conditions on $(a,\bar b)$ mentioned thus far are
necessary conditions for cancellation of some global anomalies but are not, in general,
necessary and sufficient.  In order to derive necessary and sufficient conditions we must
view anomaly cancellation from a more geometric perspective.

\section{Geometrical Anomaly Cancellation}\label{sec:GAC}

The space of all fields in a six-dimensional supergravity is fibered over the space of
nonanomalous fields
\bel{eq:10}
\fB = {\rm Met}(\CM_6) \times {\rm Conn}(\CP) \times \CS
\ee
where ${\rm Conn}(\CP)$ is the set of connections on a principal $G$ bundle $\CP \to \CM_6$ and
$\CS$ is the set of scalar fields coming from the (half-)hypermultiplets and tensormultiplets.
Path integrals in the theory, such as the partition function, can thus be written as an
integral first over the anomalous fields and second over the fields $\fB$:
\bel{eq:11}
Z_{\rm Sugra} = \int_{\fB/\fG} \int_{\rm Fermi + B} e^{-S_{\fB} - S_{\rm Fermi+B} }
\ee
where we have separated the action as a sum of an action $S_{\fB}$ involving only the nonanomalous
fields and a term $S_{\rm Fermi +B}$ giving the coupling of the anomalous fields to $\fB$.
Also,  $\fG$ is the group of diffeomorphisms and vectormultiplet gauge transformations. The problem is that
the path integral over the anomalous fields involves determinants of chiral Dirac operators and
is hence, \emph{a priori},  a section of a line bundle $\CL_{\rm Anomaly} \rightarrow \fB/\fG$:
\bel{eq:12}
\Psi_{\rm Anomaly}(\fb):=  \int_{\rm Fermi + B} e^{ - S_{\rm Fermi+B} }
\ee
where $\fb\in \fB$. Indeed, this line bundle is a determinant (or Pfaffian) line bundle constructed
from the relevant Dirac operators.
Even in the most formal sense it does not make sense to integrate a section of a line bundle against
a well-defined measure. Rather one needs to provide a \emph{geometrical trivialization} of the
line bundle $\CL_{\rm Anomaly}$. A geometrical trivialization is a (natural) choice of
 a flat connection with trivial holonomy. Cancellation of local anomalies is the flatness condition.
 Cancellation of global anomalies is the trivial holonomy condition.  Given a geometrical trivialization
 one can identify $\Psi_{\rm Anomaly}$ with a well-defined function on $\fB/\fG$, which
can then be integrated against the measure $[d\fb] e^{-S_{\fB} }$
to produce a well-defined path integral. (Of course, this path integral only makes sense in a
rather formal sense since one then needs to deal with the problems of functional integration of a
six-dimensional field theory including gravity.)
This is the geometrical formulation of anomalies that was quite popular in the 1980's.
See  \cite{AlvarezGaume:1983ig,AlvarezGaume:1984dr,Atiyah:1984tf,Freed:1986zx,Moore:1984dc,Moore:1984ws}  for accounts.

A useful modern perspective on geometrical anomaly cancellation makes use of the notion
of invertible field theory \cite{Freed:2004yc}. This was explained nicely in \cite{Freed:2014iua} and we adopt
that viewpoint here: Attached to our six-dimensional field theory will be a
\underline{seven-dimensional} invertible field theory $\CF_{\rm Anomaly}$
called the \emph{anomaly field theory} of the supergravity theory.
Here and henceforth we use the term ``field theory''   in the
mathematical sense of \cite{Atiyah:1989vu,Freed:2012hx,Freed:2016rqq,Lurie:2009keu,Segal:2002ei}
wherein a $d$-dimensional field theory
is identified with a functor $\CF$ from a bordism category of manifolds, of dimension $\leq d$, equipped
with some geometrical and/or topological structure, to some
tensor category, often the category of vector spaces. Thus, in a $d$-dimensional field
theory the partition function on a closed $d$-manifold is a complex number, the partition
function on a closed $(d-1)$-manifold is the vector space of states, and so on. In an
invertible field theory the partition function is always a nonzero complex number,
while the space of states is a one-dimensional vector space. One can put unitary structures
on invertible field theories and the deformation classes of such theories have been
classified \cite{Freed:2016rqq}.

To make contact between $\CF_{\rm Anomaly}$ and the older geometrical formulation of
anomalies we note that the anomaly field theory   of a $d$-dimensional theory is a $(d+1)$-dimensional field theory
defined on a bordism category of spin manifolds endowed with the non-anomalous fields $\fB$ in the $d$-dimensional theory
(together with their $(d+1)$-dimensional counterparts). The evaluation of an invertible $(d+1)$-dimensional
field theory on a closed $d$-manifold $\CM_d$ equipped with a point $\fb \in \fB$ is a one-dimensional vector space.
In particular, for an anomaly field theory,  $\CF_{\rm Anomaly}(\CM_d, \fb)$ is a one-dimensional vector space, but it does not
have a canonical choice of basis vector. So, varying the point $\fb$ in
$\fB$ we get a line bundle over $\fB$. This is the anomaly line bundle $\CL_{\rm Anomaly}$ of the older
geometrical formulation of anomalies.
\footnote{A useful perspective on the anomaly field theory is that it is an example of a ``field theory valued in
a field theory in one higher dimension,'' a notion discussed in \cite{MooreLectures}. This notion is closely
related to ``relative field theory,'' discussed in \cite{Freed:2012bs}.}
In the modern language of anomaly field theories, geometrical anomaly cancellation is then a two-step procedure:

\begin{enumerate}

\item Construct a ``counterterm'' invertible field theory $\CF_{\rm CT}$ which produces a  line bundle
over $\fB/\fG$ complex conjugate to $\CL_{\rm Anomaly}$,
or, equivalently, a family of one-dimensional vector spaces
$\CF_{\rm CT}(\CM_d, \fb)$ with equivariance properties under the gauge group $\fG$ on $\fB$ conjugate to
those of the family $\CF_{\rm Anomaly}(\CM_d, \fb)$.

\item Use the data of the $d$-dimensional fields to give a \underline{local} construction of a section
$\Psi_{\rm CT} \in \CF_{\rm Anomaly}(\CM_d)$ so that
\bel{eq:13}
\int_{\rm Fermi+B} e^{-S_{\rm Fermi+B}} \Psi_{\rm CT}
\ee
descends to an honest function on $\fB/\fG$.

\end{enumerate}

One might well ask: Why not simply take $\CF_{\rm CT} = \CF_{\rm Anomaly}^* $ ?  This would certainly
trivially cancel the anomalies! The point is that one must then also construct a vector in
$\CF_{\rm Anomaly}^*(\CM_d,\fb)$ using just the fields of the $d$-dimensional field theory in a way which
is local. In older language, one must construct a local counterterm. That is the real challenge.

The anomaly field theory $\CF_{\rm Anomaly}$  of a six-dimensional supergravity based on
\bel{eq:data}
(G,\CR,\Lambda,a,\bar b)
\ee
is a product of 7-dimensional Dai-Freed field theories \cite{Dai:1994kq} associated with the chiral fermions and
(anti-) self-dual fields. (This is essentially a restatement of the formulae for global anomalies derived
by Witten in \cite{Witten:1985xe,Witten:1999eg}.) To define the Dai-Freed theory
recall   that if $D$ is a Dirac operator on odd-dimensional manifolds then the
APS eta invariant:
\bel{eq:14}
\xi(D) := \half ( \eta(D) + \dim\ker(D))
\ee
defines an invertible field theory on the category of Riemannian spin manifolds equipped with suitable gauge bundle
and connection. Indeed $\CF_{\rm Dai-Freed}(\CM_{2n+1})= e^{2\pi \I \xi(D)}$ is the nonzero partition function
while the same expression $e^{2\pi \I \xi(D)}$ on $2n$-dimensional manifolds defines a section of a nontrivial
line bundle over the space of equivalence classes of metrics and connections.
\footnote{The $\eta$ invariant is closely
related to the Chern-Simons invariant. It is well-known that the Chern-Simons invariant on a $(2n+1)$-manifold
with $2n$-dimensional boundary is not gauge invariant. This failure of gauge invariance can be interpreted as the
statement that the exponentiated Chern-Simons invariant is a section of a line bundle over the space of gauge
equivalence classes of connections. In a similar way the exponentiated $\eta$ invariant should be interpreted
as a section of a line bundle.}
In fact, that line bundle is again a determinant line bundle of a six-dimensional chiral Dirac operator.
See \cite{Witten:2015aba} for an account in the physics literature.

It is shown in \cite{MonnierMoore} how to construct a Dirac operator $D$ from the data $(G,\CR,\Lambda)$
so that the anomaly field theory for the six-dimensional supergravity is the Dai-Freed field theory for $D$.
\footnote{The contribution of the (anti-)self-dual fields involves some tricky factors of two.
It is really a ``quarter Dai-Freed theory.''}
It is useful to examine the partition function of this seven-dimensional invertible field theory on
\underline{closed} seven-dimensional spin manifolds $\CU_7$ equipped with $\fb\in \fB$.
In general, $\eta$ invariants are impossible to compute in simpler terms. However, if the principal
$G$-bundle $\CP \to \CU_7$ extends to a principal $G$-bundle over an eight-dimensional spin manifold $\CW_8$,
with $\p \CW_8 = \CU_7$, and if we choose anomaly coefficients $(a,\bar b)$   so that $I_8 = \half Y^2$
then a rather simple expression emerges \cite{MonnierMoore}:
\bel{eq:15}
\CF_{\rm Anomaly}(\CU_7,\fb) = \exp\left\{ 2\pi \I \left( \int_{\CW_8} \half Y^2 - \frac{\sigma}{8} \right)\right\}
\ee
where $\sigma$ denotes the signature of the free quotient $H^{4,free}(\CW_8,\p \CW_8; \Lambda)$ of the relative cohomology group.
Since the signature is multiplicative this is the product of the signature $\sigma(\Lambda)=(1-T)$ of $\Lambda$ and the signature
of $H^{4,free}(\CW_8,\p \CW_8; \IZ)$.
As we will see momentarily,  equation \eqref{eq:15}  provides the key to the construction of the counterterm invertible field theory.

The simple expression \eqref{eq:15} raises the question of whether there can be topological obstructions
to extending $\CP \to \CU_7$ to a bounding spin manifold. Indeed such obstructions can exist.
The bordism group $\Omega_7^{\rm spin}(pt)=0$,
meaning that every closed spin seven-dimensional manifold is a boundary of some spin eight-dimensional manifold. However,
if we wish to extend a principal $G$ bundle to 8 dimensions then the appropriate bordism group is
 $\Omega_7^{\rm spin}(BG)$, and this group can be nonzero. For example, if $G=O(n)$ then the integral
 $\int_{\CU_7} w_1(\CP)^7 $ is a bordism invariant. It would vanish if the principal $O(n)$ bundle $\CP$ extended
 to $\CW_8$. But it is easy to construct examples where this integral is nonzero. Nevertheless, it is shown in
 \cite{MonnierMoore} that $\Omega_7^{\rm spin}(BG)=0$ for many groups of interest, including $G=U(n), SU(n), USp(2n)$
 and products thereof. (It can also be shown to vanish for some special cases, such as $G=E_8$.)
  On the other hand, $\Omega_7^{\rm spin}(BG)$ is nonzero and computable for all nontrivial cyclic groups.

We now make an elementary algebraic manipulation of uncompleting the square.
We define $X := Y-\half \lambda'$ where $\lambda' := a \otimes \lambda''$ and
$\lambda''$ is, roughly speaking,
 $\lambda'' \sim \half p_1$. Then  $X$ is a closed 4-form with periods in $\Lambda$ (thanks to \eqref{eq:9})
 and \eqref{eq:15} takes the form
\bel{eq:16}
\CF_{\rm Anomaly}(\CU_7,\fb) = \exp\left\{ 2\pi \I \left( \int_{\CW_8} \half X (X+\lambda') + \frac{\lambda'^2-\sigma}{8}  \right)\right\} \, .
\ee
Here $\lambda'^2$ is the scalar given by the natural pairing obtained by multiplying, contracting with the metric on $\Lambda$, and
integrating over $\CW_8$.

(The reason for the qualifier ``roughly speaking'' in the previous paragraph is that there
is a technical, but quite important subtlety.  It leads to much of the hard work in
\cite{MonnierMoore} and we only briefly describe it here. Some readers will wish to skip this paragraph.
When $\CW_8$ is a compact spin $8$-manifold without boundary there is a canonical cohomology class $\lambda \in H^4(\CW_8;\IZ)$
such that $p_1(\CW_8)=2\lambda$. The class $\lambda$
 descends to a characteristic vector for the unimodular lattice $H^{4,free}(\CW_8;\IZ)$.
Happily, such a characteristic vector satisfies $\int_{\CW_8} \lambda^2 = \sigma(H^{4,free}(\CW_8;\IZ)) \mod 8$.
Finally, $\lambda$ is an integral lift of the fourth Wu class of $\CW_8$.
However, in our application $\CW_8$ is a manifold with nontrivial boundary $\CU_7$, so $H^{4,free}(\CW_8;\IZ)$,
and more to the point, the relative cohomology group $H^{4,free}(\CW_8, \p \CW_8;\IZ)$ are not unimodular and
$\lambda$ is no longer a characteristic vector.
What we must do is choose a smooth relative cocycle $\lambda''$ which, when reduced modulo two represents
the fourth Wu class of $\CW_8$ and yet vanishes on the boundary. This is possible since the fourth Wu class
vanishes on closed manifolds of dimension lower than $8$. The relative cocycle $\lambda''$ is of the form $d\eta + \hat \nu$,
where $\hat \nu$ is an integral lift of the Wu class and $d\eta$ trivializes the Wu class on the boundary
$\CU_{7}$. We then define $\lambda' := a\otimes \lambda'' $. If $\CW_8$ were closed then, since
$a^2 = \sigma(\Lambda) \mod 8$ by the factorization conditions, the ``extra'' eighth root of unity in the second
term on the right hand side of \eqref{eq:16} could be dropped.  But since
$\p \CW_8 = \CU_7$ is nonempty the extra phase is nontrivial and cannot be dropped.
 It will lead to a related technicality in our construction of the counterterm invertible field theory.)

\section{A Construction Of The Counterterm Invertible Field Theory From Wu Chern-Simons Theory}

The importance of \eqref{eq:16} is that it is also closely related to a formula for both
the exponentiated action and the partition function of a
7-dimensional analog of Abelian spin-Chern-Simons theory known as Wu Chern-Simons (WCS) theory.
We will use
a modified version of the WCS partition function to construct the counterterm invertible field theory $\CF_{\rm CT}$.

A general theory of Wu Chern-Simons theories has been presented in \cite{Monnier:2016jlo} and the detailed
computations of \cite{MonnierMoore} make extensive use of the results of \cite{Monnier:2016jlo}.
In the case of six-dimensional supergravity we need the seven-dimensional WCS theory of a
$3$-form gauge potential with $4$-form fieldstrength $X \in \Omega^4(\CU_7;\Lambda)$.
The analogy with $3$-dimensional spin Chern-Simons theory where the gauge group is a torus $\Lambda_{\IR}/\Lambda$
and $\Lambda$ is an integral lattice is quite illuminating. (See, for example \cite{Belov:2005ze} for a
detailed discussion of these theories.) In the latter theories, if we normalize the fieldstrength $F \in \Omega^2(\CU_3;\Lambda_{\IR})$
so that its periods are in $\Lambda$ then the action is
\bel{eq:3dSCS}
 \exp\left\{ 2\pi \I \left( \int_{\CW_4} \half F (F+\lambda') \right)\right\} \, .
\ee
Here $\lambda' = W \otimes \hat w_2$, while $W$ is a characteristic vector of $\Lambda$,
\footnote{In general, if $\Lambda$ is an integral lattice then $v^2 \mod 2$ is a linear
function on vectors $v\in \Lambda$. A  characteristic vector
$W\in \Lambda^\vee$ is a vector such that $v^2 = v \cdot W \mod 2$, thus making
the obvious manifest.}
$\hat w_2$ is an integral lift of the second Stiefel-Whitney class, and the fields have been
extended to a bounding four-manfiold $\CW_4$ (again so that $F$ has periods in $\Lambda$).
 Because \eqref{eq:3dSCS} involves an integral
over a manifold with boundary we need to take into account the boundary behavior of the
integrand. In particular, it is necessary to choose a trivialization of   $\hat w_2$ on the boundary.
Such a trivialization amounts to a choice of spin structure on $\CU_3$, and the  action
\eqref{eq:3dSCS} depends on the choice of spin structure.

Just as the Abelian spin Chern-Simons theory in three dimensions depends on a choice of
spin structure, the  construction of the $7$-dimensional Wu Chern-Simons
 theory requires a choice of \emph{Wu structure} $\omega$.
Wu structures are higher-form generalizations of spin structures. In our case a
Wu structure is a trivialization of the fourth Wu class. (Since our manifolds are
orientable and spin the fourth Wu class coincides with the fourth Stiefel-Whitney class.)
The isomorphism classes of Wu structures on $\CU_7$ or $\CM_6$  form a torsor for
$H^3(\CU_7; \IZ_2)$ or $H^3(\CM_6; \IZ_2)$, respectively. In addition, in complete analogy
to the three-dimensional case,  the formulation of the
Wu Chern-Simons theory requires a choice of a characteristic vector $\tilde a \in \Lambda^\vee$ \cite{Monnier:2016jlo}.
With these choices the action of the WCS theory on $\CU_7$ can be written as:
\bel{eq:17}
\CF_{\rm WCS}^{\rm PQ}(\CU_7;\omega; X) = \exp \left\{ 2\pi \I \int_{\CW_8} \half X (X + \tilde \lambda) \right\}
\ee
where the superscript PQ (for ``pre-quantum'')  indicates that we are defining an invertible field theory from the
\underline{classical action} of the WCS theory. On the RHS of \eqref{eq:17} we have
 $\tilde \lambda = \tilde a \otimes \lambda''$. Again, roughly speaking, $\lambda'' \sim \half p_1$. More
 accurately, it is described in the final paragraph of Section \ref{sec:GAC}, and it depends on a choice of
 Wu structure.
There is no topological obstruction to extending the field $X$, initially defined on $\CU_7$, to
$\CW_8$. However, it is critical that $\tilde a$ be a characteristic vector of $\Lambda$ in order for the action
to be independent of extension to eight dimensions.

Comparing equations \eqref{eq:16} and \eqref{eq:17} it is evident that we should try to use
$\CF_{\rm WCS}^{\rm PQ}$ to define the counterterm invertible field theory. This is not exactly
the invertible field theory we need because we must
take into account the ``extra'' eighth root of unity appearing in \eqref{eq:16}.
Rather, what we do instead is
 regard $X$ as a background, nondynamical field coupled to a dynamical flat 4-form
$Z \in \Omega^4(\CU_7;\Lambda_{\IR})$ (or, to be more accurate, a flat differential
4-cocycle with coefficients in $\Lambda$ - see below). The partition function of the $Z$-theory, where we integrate
over all flat $4$-forms, depends on
$X$ (and Wu structure) and differs from the action $\CF_{\rm WCS}^{\rm PQ}$ by
the Arf invariant of a certain quadratic refinement $q$ of the link pairing of torsion classes in $H^4(\CU_7;\Lambda)$.
\footnote{The quadratic refinement $q$ is itself defined in terms of a Wu Chern-Simons action $q(z) =S(Z)$
where $Z$ is a field with torsion topological class $z$.}
This partition function satisfies gluing rules so it is part of a 7-dimensional field theory
which we call the \emph{Wu Chern-Simons theory}. To pursue the analogy with Abelian spin Chern-Simons
theory, $X$ plays the role of the external Maxwell electromagnetic field and the Wu Chern-Simons theory is
analogous to the effective action for the Maxwell field after integrating out the statistical Chern-Simons
fields used when modeling the FQHE.  Returning to our Wu Chern-Simons field theory, its value on a $7$-manifold $\CU_7$ endowed with a Wu
structure and with a background field  $X$ is
\bel{eq:WCS}
\CF_{\rm WCS}(\CU_7;\omega;X) = \exp \left\{ 2\pi \I \int_{\CW_8} \half X (X + \tilde \lambda) \right\} {\rm Arf}(q) \, .
\ee
It turns out that the phase ${\rm Arf}(q)$ is
\bel{eq:ExtraPhase}
\exp\left\{ 2\pi \I \left(  \frac{\tilde\lambda^2-\sigma}{8}  \right)\right\}
\ee
and is thus exactly of the same form as the ``extra'' phase in \eqref{eq:16}.
Moreover, when we construct $\CF_{\rm WCS}$ as a field theory then it is only an
\underline{invertible} field theory when $\Lambda$ is unimodular.

We are now ready to construct the counterterm theory. We take $\tilde a = a$. This
 makes sense when $\Lambda$ is unimodular and identifies $\tilde \lambda = \lambda'$.
  Next, we take the background field to be  $X = Y(\fb)- \half \lambda'$,
where we write $Y(\fb)$ rather than $Y$ to emphasize the dependence on the nonanomalous fields $\fb\in \fB$,
as in equation \eqref{eq:6}.
Finally, since we want to cancel anomalies, we take a complex conjugate.  As we have
stressed, the quantities in  \eqref{eq:17} and \eqref{eq:WCS} depend on a choice of Wu structure $\omega$.
We do not wish to add a choice of Wu structure to the defining data of a supergravity theory!
(And, more rationally, the bordism category on which $\CF_{\rm Anomaly}$ is defined does not include
a choice of Wu structure.)  Recall that  $\lambda''$ is $\omega$-dependent. In \cite{MonnierMoore} we show that
the dependence can be chosen so that the choice of Wu structure actually cancels out
when we evaluate
\bel{eq:DropOut}
\CF_{\rm WCS}(\CU_7;\omega;X = Y(\fb)- \half \lambda')
\ee
(and similarly if we substitute $\CU_7 \to \CM_6$).
Since $Y(\fb)$ (or rather, its lift to a differential cocycle - see below) depends on the metric and vectormultiplet
gauge connection the counterterm field theory
\bel{eq:18}
\CF_{\rm CT}(\CM_6; \fb):= \left( \CF_{\rm WCS}(\CM_6;  X=Y(\fb)-\half \lambda' ) \right)^*
\ee
is defined on the same geometrical category as $\CF_{\rm Anomaly}$. This is our counterterm
invertible field theory.

We now consider the product theory:
\bel{eq:19}
\CF_{\rm Top}:= \CF_{\rm Anomaly} \times \CF_{\rm CT}  \, .
\ee
It follows from equation \eqref{eq:16} and equation \eqref{eq:17} that
the  dependence on metric and gauge connection cancel out in this invertible field theory and
we are therefore left with a \underline{spin topological field theory}. This theory is determined by
its seven-dimensional partition function. By construction, the partition function is trivial
when the principal bundle $\CP \to \CU_7$ extends to $8$ dimensions, and therefore the
partition function is a homomorphism $\Omega_7^{\rm spin}(BG) \to U(1)$.
If this homomorphism is nontrivial we can be confident that the Green-Schwarz mechanism, at least insofar
as we define it here,  cannot cancel the global gravitational anomalies.

 While it is in general impossible
to evaluate eta invariants in simpler terms, one can show, using results from
\cite{Gilkey},  that for certain $G$-bundles over Lens spaces
$\CF_{\rm Top}$ will in general be nontrivial when $G$ is a cyclic group.  On the other hand, if the
theory $\CF_{\rm Top}$ is the trivial theory we can proceed to the next step in anomaly cancellation,
namely a local construction of a vector in the statespace $\CF_{\rm CT}(\CM_6; \fb)$ where $\fb\in \fB$.
Note that if $\Omega_7^{\rm spin}(BG) = 0 $ then  $\CF_{\rm Top}$ is necessarily trivial. Moreover,
when $\Omega_6^{\rm spin}(BG) \not= 0  $ there can be different ``settings'' \cite{Freed:2004yc} of the integrand of the
supergravity path integral. This leads to new $\theta$-like angles, which do not seem to have been discussed in
the literature on 6d anomaly cancellation.
 (See also pp. 19-20 of \cite{Witten:1996hc} where a similar phenomenon was forseen in a different
context.)

We remark that it is significant that our construction requires $\Lambda$ to be unimodular.
This condition has been previously derived in \cite{SeibergTaylor} by consideration of global anomalies.
We can consider the present discussion as an independent derivation that $\Lambda$ is self-dual. Moreover we
have found that  $a=\tilde a$ must be a characteristic vector of $\Lambda$. While it has been found
that in all $F$-theory constructions $a$ must be a characteristic vector, no derivation of this
condition within the context of low energy supergravity has previously been given. We can regard
the present discussion as a low-energy demonstration that $a$ must be a characteristic vector.

\section{Differential Cohomology And The Green-Schwarz Term}

In order to make precise sense of several of the claims made above, and, especially, to give an
explicit local construction of the Green-Schwarz term in topologically nontrivial situations
one must make use of the formalism of differential cohomology. Differential cohomology is a
mathematical framework for describing higher form abelian gauge fields (and global symmetries).
See \cite{springerlink:10.1007/BFb0075216,Dijkgraaf:1989pz,Freed:2000ta,Freed:2006ya,Freed:2006yc,hopkins-2005-70} for expositions.

Briefly, associated to a $p$-form Abelian gauge field $A$ are three distinct
gauge invariant quantities. There is, of course, the fieldstrength, (or curvature), $F=dA$.
Moreover, there are ``Wilson lines'' (or holonomies), $\exp( 2\pi\I \int_{\Sigma} A)$ where $\Sigma$ is a
$p$-cycle and $A$ is normalized so that $F$ has integer periods. Finally, there is the
topological class, or characteristic class, which is valued in
$E^*$, where $E^*$ is some generalized cohomology theory. The gauge equivalence class of a
$p$-form gauge field is an element of a differential cohomology group associated to $E^*$.  The
differential cohomology group is an
infinite dimensional Abelian group and the relation of this group to the above three gauge
invariant quantities is summarized by some standard exact sequences.

In our case, we have modeled the $B$ field using the generalized cohomology theory $E^*(\cdot) = H^*(\cdot; \Lambda)$.
In order to write a precise form of the Green-Schwarz term $\Psi_{\rm CT}$ generalizing \eqref{eq:4}
to topologically nontrivial situations we need to lift $Y\in \Omega^4(\CM_6;\Lambda)$ to a
differential cocycle. Given a metric, spin structure, and vectormultiplet gauge connection there is
a way  of lifting $Y$ to a differential
cocycle $\check Y$.
\footnote{The procedure is described in detail in Appendix A of \cite{MonnierMoore} and involves
making some universal choices on classifying spaces.}
The lift is not quite canonical, and while all lifts $\check Y$ have the same fieldstrength (by definition)
they can differ by exact differential cocycles and, more importantly, they can differ in their torsion components.
While these torsion components are  not visible in the fieldstrength $Y$ specified in \eqref{eq:6},  they
do affect global anomaly cancellation. The anomaly coefficient of $\check Y$ associated to the vectormultiplet
gauge symmetry is really an element $b \in H^4(BG;\Lambda)$. The projection to the free quotient is the quantity
$\half \bar b \in H^4(BG_1;\Lambda)$ discussed above. Any lift of $\half \bar b$  to
$H^4(BG;\Lambda)$ yields an acceptable set of anomaly coefficients. Any lift that cancels global anomalies
(if it exists) is physically acceptable. Given any reference lift, we can encode the torsion component of the
anomaly coefficients by an element
\bel{eq:20}
b_T \in  {\rm Tors}(H^4(BG;\Lambda)) \, .
\ee
$b_T$ is a new piece of data that must be added to the defining data of a
low energy six-dimensional supergravity theory. Hence we arrive at a new necessary condition
for global anomaly cancellation: A supergravity theory based on $(G,\CR,\Lambda, a,  b)$
can only be nonanamolous if the topological field theory defined in \eqref{eq:19} is trivial.

In section 6 of \cite{MonnierMoore}  we  use the formalism of differential
cohomology to construct an explicit Green-Schwarz counterterm $\Psi_{\rm CT}$
which can be seen as a section of the counterterm line bundle over $\fB$.
 The construction is independent of any choice of Wu structure, is
completely local in the six-dimensional fields (yet makes sense in nontrivial topology),
and of course it reduces to the standard Green-Schwarz counterterm in topologically trivial situations.
When the topological field theory \eqref{eq:19} is trivial this counterterm cancels
all local and global anomalies. Thus, the triviality of \eqref{eq:19} is both necessary and sufficient.

\section{The Main Result, And A Corollary}

In summary, the main result of \cite{MonnierMoore} is the
following theorem:   If a supergravity theory is defined by the data $(G,\CR,\Lambda, a,  b)$ such that
\begin{enumerate}
\item $I_8 = \frac{1}{2} Y^2$;
\item $\Lambda$ is unimodular and $a\in \Lambda$ is a characteristic vector;
\item $\bar b \in 2 H^4(BG_1;\Lambda)$;
\end{enumerate}
then a necessary and sufficient condition for the cancellation of all global anomalies is that
the topological field theory \eqref{eq:19} is trivial. This reduces to the condition that
the partition function  on 7-manifolds, which is a homomorphism  $\Omega^{\rm Spin}_7(BG) \rightarrow U(1)$,
must be trivial.

An important corollary is that if the above three criteria are met and $G$ is such that
 $\Omega^{\rm Spin}_7(BG) = 0$ then all anomalies, both local and global, are
 cancelled by the Green-Schwarz mechanism.

We must close with a word of caution. In our work we have made what we consider to be the most
reasonable choice of the generalized cohomology theory used to model the (anti-) self-dual fields of
six-dimensional supergravity. We would note, however, that other generalized cohomology theories are
used for anomaly cancellation in type II strings and orientifolds
\cite{Distler:2009ri,Distler:2010an,Freed:2000ta,Minasian:1997mm}.
It would be very interesting if the adoption of a different generalized cohomology theory for
the $B$-fields of six-dimensional supergravity led to different conditions for a nonanamolous
field content $(G,\CR,\Lambda)$ of six-dimensional supergravity.

\subsection*{Acknowledgements}

We would like to thank Daniel Park for discussions that led to this project. We also thank Dan Freed, Mike Hopkins, Graeme Segal,
Nati Seiberg,
Wati Taylor, Andrew Turner and
Edward Witten for useful discussions. G.M. is supported by the DOE under grant DOE-SC0010008 to Rutgers University. S.M. is supported in part by the grant MODFLAT of the European Research Council, SNSF grants No. 152812, 165666, and by NCCR SwissMAP, funded by the Swiss National Science Foundation. Finally, GM would like to thank the organizers of the Strings 2018 conference for the invitation to speak.

\end{document}